\documentclass{article}

\usepackage{arxiv}

\usepackage[utf8]{inputenc} % allow utf-8 input
\usepackage[T1]{fontenc}    % use 8-bit T1 fonts
\usepackage{hyperref}       % hyperlinks
\usepackage{url}            % simple URL typesetting
\usepackage{booktabs}       % professional-quality tables
\usepackage{amsfonts}       % blackboard math symbols
\usepackage{nicefrac}       % compact symbols for 1/2, etc.
\usepackage{microtype}      % microtypography

\usepackage{lipsum}         % Can be removed after putting your text content
\usepackage{graphicx}
\usepackage{natbib}
\usepackage{doi}

\usepackage{booktabs}
\usepackage{threeparttable}
\usepackage{tabularx}
\usepackage{geometry}
\usepackage{multirow}
\usepackage{amsmath} 

\usepackage{cleveref}       % smart cross-referencing

\title{DDEvENet: Evidence-based Ensemble Learning for Uncertainty-aware Brain Parcellation Using Diffusion MRI}

\date{}

\usepackage{authblk}
\author[1]{Chenjun Li\thanks{Chenjun Li and Dian Yang are co-first authors.}}
\author[1]{Dian Yang$^*$}
\author[2]{Shun Yao}
\author[3]{Shuyue Wang}
\author[4]{Ye Wu}
\author[1]{Le Zhang}
\author[5]{Qiannuo Li}
\author[6]{Kang Ik Kevin Cho}
\author[6]{Johanna Seitz-Holland}
\author[6]{Lipeng Ning}
\author[6]{Jon Haitz Legarreta}
\author[6]{Yogesh Rathi}
\author[6]{Carl-Fredrik Westin}
\author[6]{Lauren J. O'Donnell}
\author[7]{Nir A. Sochen}
\author[6]{Ofer Pasternak$^\dagger$}
\author[1]{Fan Zhang\thanks{Fan Zhang and Ofer Pasternak are co-senior authors, Fan Zhang is the corresponding author.}}

\affil[1]{University of Electronic Science and Technology of China, Chengdu, Sichuan, China}
\affil[2]{The First Affiliated Hospital, Sun Yat-sen University, Guangzhou, Guangdong, China}
\affil[3]{The Second Affiliated Hospital, Zhejiang University School of Medicine, Hangzhou, Zhejiang, China}
\affil[4]{Nanjing University of Science and Technology, Nanjing, Jiangsu, China}
\affil[5]{East China University of Science and Technology, Shanghai, China}
\affil[6]{Harvard Medical School, Boston, MA, USA}
\affil[7]{School of Mathematical Sciences, University of Tel Aviv, Tel Aviv, Israel}

\renewcommand{\shorttitle}{\textit{arXiv} Template}

\hypersetup{
pdftitle={A template for the arxiv style},
pdfsubject={q-bio.NC, q-bio.QM},
pdfauthor={David S.~Hippocampus, Elias D.~Striatum},
pdfkeywords={First keyword, Second keyword, More},
}

\begin{document}
\maketitle

\begin{abstract}
	 In this study, we developed an Evidence-based Ensemble Neural Network, namely EVENet, for anatomical brain parcellation using diffusion MRI. The key innovation of EVENet is the design of an evidential deep learning framework to quantify predictive uncertainty at each voxel during a single inference. To do so, we design an evidence-based ensemble learning framework for uncertainty-aware parcellation to leverage the multiple dMRI parameters derived from diffusion MRI. Using EVENet, we obtained accurate parcellation and uncertainty estimates across different datasets from healthy and clinical populations and with different imaging acquisitions. The overall network includes five parallel subnetworks, where each is dedicated to learning the FreeSurfer parcellation for a certain diffusion MRI parameter. An evidence-based ensemble methodology is then proposed to fuse the individual outputs. We perform experimental evaluations on large-scale datasets from multiple imaging sources, including high-quality diffusion MRI data from healthy adults and clinically diffusion MRI data from participants with various brain diseases (schizophrenia, bipolar disorder, attention-deficit/hyperactivity disorder, Parkinson’s disease, cerebral small vessel disease, and neurosurgical patients with brain tumors). Compared to several state-of-the-art methods, our experimental results demonstrate highly improved parcellation accuracy across the multiple testing datasets despite the differences in dMRI acquisition protocols and health conditions. Furthermore, thanks to the uncertainty estimation, our EVENet approach demonstrates a good ability to detect abnormal brain regions in patients with lesions, enhancing the interpretability and reliability of the segmentation results.
\end{abstract}

% keywords can be removed


\section{Introduction}
Parcellation of cortical and subcortical brain regions is a vital step in neuroimaging analysis for mapping the structural and functional regions of the brain \citep{JI201935,10.3389/fnana.2012.00034}. Diffusion MRI (dMRI) \citep{BASSER2011560} is an advanced MRI technique that characterizes tissue microstructure and uniquely enables in vivo mapping of the brain white matter connections \citep{ZHANG2022118870}. Parcellation is an essential step for many computational dMRI applications, such as fiber tract identification \citep{Wassermann2016}, structural connectome construction \citep{sporns2005human}, and clinical applications, such as characterizing abnormal regional microstructure across brain disorders \citep{seitz2018alteration}. Most approaches for dMRI parcellation compute the parcellation from anatomical MRI (T1- or T2-weighted) data and then register it to the dMRI space. However, the inter-modality registration presents difficulties due to image distortions \citep{albi2018image,wu2008comparison,jones2010twenty},  and the low resolution of dMRI data \citep{malinsky2013registration}, often leading to inaccuracies in the resulting brain parcellation. Additionally, these methods cannot be applied without collecting accompanying anatomical MRI data. 

\begin{figure}[htbp]
  \centering
  \includegraphics[width=0.8\textwidth]{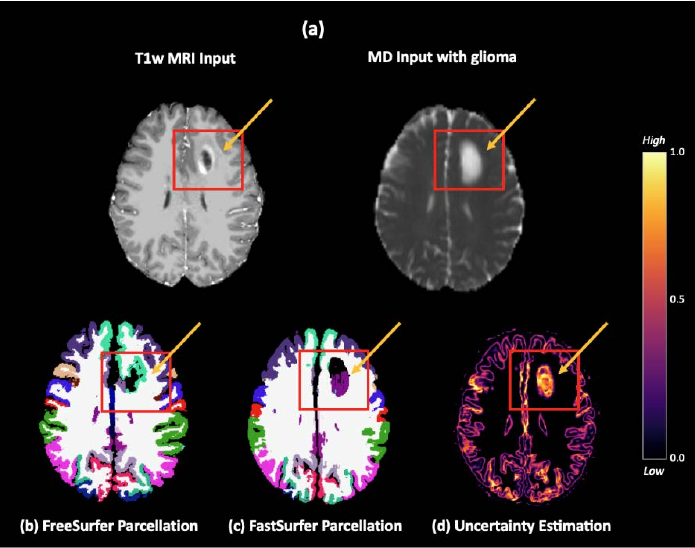}
  \caption{Example of out-of-distribution parcellation. (a) shows the input image data with apparent abnormal lesion regions (glioma) within the red rectangle, which constitutes this scan as an out-of-distribution scan. (b) shows the parcellation result using the widely used FreeSurfer software \citep{fischl2012freesurfer}, where the lesion boundary is mislabeled as being part of the cortex. (c) shows the parcellation result using FastSurfer \citep{henschel2020fastsurfer}, a recently proposed deep-learning approach. While we can observe a visually improved parcellation, there is still apparent mislabeling in the glioma (shown with yellow arrows). (d) shows the uncertainty estimation using our proposed EVENet method, which outlines tissue boundaries as well as abnormal brain regions and is utilized to improve parcellation. 
}
  \label{FIG.1}
\end{figure}

Advanced techniques have been developed for direct brain parcellation from dMRI data using deep learning, eliminating the need for inter-modality registration and thus enhancing parcellation performance \citep{liu2007brain,wen2013brain,yap2015brain,ciritsis2018automated,zhang2021deep,theaud2022doris,zhang2023ddparcel}, with improved accuracy and efficiency \citep{theaud2022doris,zhang2023ddparcel,billot2023synthseg}. Early work using deep learning typically employs encoder-decoder architecture with convolutional neural networks (CNNs) for efficient feature extraction and segmentation \citep{zhang2021deep,theaud2022doris}. More recent methods \citep{li2023can,hayat2024transformer} have considered the use of data fusion techniques with attention mechanisms to leverage more contextual information and improve accuracy. However, an important limitation of these advanced methods is that they were trained and tested on data that are closely matched in terms of subject characteristics and imaging acquisition protocols. Their effectiveness diminishes when applied to out-of-distribution data, i.e., data that is different from the training data. As a result, the generalizability of the existing dMRI parcellation methods to data from different sources remains a challenge.  In general, there are several major sources of variability that lead to uncertainty and poorer generalizability in deep learning model predictions in dMRI parcellation. First, it is well known that variations across acquisition sites, arising from different MRI scanners and/or acquisition protocols can result in significant biases in dMRI data \citep{karayumak2019retrospective,tax2019cross,li2024diffusion}.. Site variability can result in poor generalizability to data from different centers/sites/scanners. Second, the parcellation performance can be compounded by differences in spatial resolution, leading to varying overlaps or partial volumes between anatomical structures \citep{guevara2020superficial}. Consequently, the certainty in the parcellation prediction in the boundary between regions or tissue types diminishes. Third, current parcellation methods are primarily designed and/or trained on the brains of healthy individuals with normal structural appearances. As a result, they can be erroneous when applied to brains with abnormalities such as lesions,  brain tumors, and white matter hyperintensity (WMH). Parcellation of out-of-distribution data (e.g., Figure \ref{FIG.1} with a dataset containing a glioma) is challenging not just for deep learning approaches but also for traditional parcellation tools such as FreeSurfer \citep{fischl2012freesurfer}.

Including uncertainty estimation techniques in the learning process is an important way to improve generalizability of deep neural networks \citep{kendall2017uncertainties,abdar2021review,zou2023review}, and is especially useful to their application on out-of-distribution data. In the context of anatomical brain parcellation, the ability to estimate uncertainty in the model's predictions can help pinpoint potential areas in which the model is not accurate, and offer extra insights into anatomical variations that cause the uncertainty. In turn, the estimation of uncertainty can be included in the training process to reduce mis-labeling in areas of high uncertainty. For example, the estimation of uncertainty for the glioma case (Figure \ref{FIG.1}) highlights lesion regions as having high uncertainty. In the literature, one widely used approach for uncertainty estimation is based on Bayesian neural networks (BNNs) \citep{springenberg2016bayesian} that calculate a probability distribution over network weights to model parameter uncertainty in each voxel. However, BNNs suffer from high computational complexity because of the difficulty of efficiently computing posterior distributions over the model parameters. Monte Carlo Dropout \citep{gal2016dropout} is a method to obtain uncertainty estimates by performing dropout at inference time and running the model multiple times. It introduces randomness into the model to let it behave like an ensemble of networks but requires careful and complicated adjustment of NN structures, which may in turn affect model performance. The Deep Ensemble approach \citep{lakshminarayanan2017simple} attempts to overcome overfitting and improves predictive performance by training multiple deep models and aggregating their predictions. dMRI data is multi-dimensional and orientation-dependent \citep{johansen2013diffusion,shi2012altered,wee2011enriched}, from which multiple parameters (e.g., fractional anisotropy; FA, and mean diffusivity; MD) can be derived to describe multiple different microstructural properties. In this case, the dMRI data is inherently suitable for the application of ensemble learning techniques \citep{winzeck2019ensemble,lella2021ensemble}, that can leverage information across different features to improve parcellation results. Yet, utilizing these ensembles necessitates model optimization for each network in the ensemble, demanding significant amounts of computational resources. 

Evidential deep learning \citep{sensoy2018evidential,zou2022tbrats,li2023region}, rooted in the principles of subjective logic theory \citep{jsang2018subjective}, offers a simple yet effective tool for uncertainty estimation. Importantly, this relatively new approach provides improved results, especially for the classification of out-of-distribution samples \citep{sensoy2018evidential}. So far, evidential deep learning has been used for few-class segmentation such as binary tumor segmentation \citep{zou2022tbrats,li2023region}, but holds great potential for multi-class segmentation (e.g. parcellation of the brain image into over tens of regions) for its high efficiency and ease of implementation. Subjective logic theory extends traditional probability theory by introducing the notion of uncertainty that integrates evidence and belief \citep{jsang2018subjective}. Unlike standard logic, where propositions are dichotomously considered true or false, or probabilistic logic, which assigns probabilities in the range [0, 1] to all arguments, evidential deep learning explicitly acknowledges a probability for uncertainty \citep{sensoy2018evidential}. In this framework, the network is trained to simultaneously optimize data fit while maximizing evidence, hence minimizing uncertainty. 

In this study, we propose the Evidence-based Ensemble Neural Network (EVENet), a novel uncertainty-aware deep learning method for anatomical brain parcellation of cortical and subcortical regions directly from dMRI data. The novelty of our approach lies primarily in the application of evidential deep learning combined with ensemble methods for uncertainty-aware brain parcellation. Our method is built on established neural network architectures and techniques but tailored to consider the multi-parameter property of dMRI data, enhancing both the accuracy and interpretability of parcellation results. The key innovation of EVENet is its utilization of evidential deep learning to quantify uncertainty at each voxel during a single inference. As a result, even though EVENet is trained using data from healthy individuals scanned with a single MRI protocol, it is able to provide accurate parcellation across different datasets with different imaging acquisitions, and from different populations, including individuals with brain disorders. In addition, we combine the uncertainty estimation across different channels, this way EVENet leverages multiple dMRI parameters to perform fine-scale parcellation, followed by an evidence-based ensemble fusion of the output from each dMRI parameter for final parcellation predictions and uncertainty estimation. Given that different diffusion parameters describe tissue properties from varying perspectives, incorporating uncertainty estimation can enhance prediction results by accounting for the differences in outputs from each parameter. We perform experimental evaluations on large-scale datasets from multiple imaging sources, including high-quality dMRI images from healthy adults and clinically dMRI data from participants with various brain disorders (including schizophrenia, bipolar disorder, attention-deficit/hyperactivity disorder, Parkinson’s disease, cerebral small vessel disease, and neurosurgical patients with brain tumors). Our results indicate not only improved parcellation accuracy but also the ability to detect abnormal brain regions, suggesting EVENet's robustness and generalization capabilities in both research and clinical settings.

\section{Methods}

\begin{figure}[htbp]
  \centering
  \includegraphics[width=1.0\textwidth]{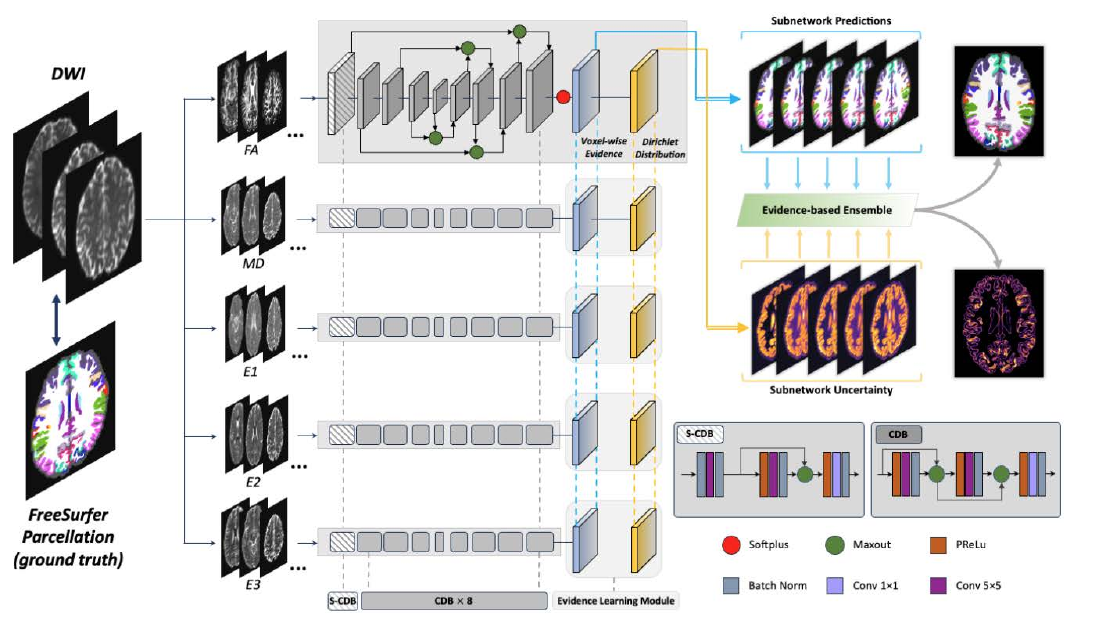}
  \caption{EVENet overview. Five parameter maps calculated from the diffusion-weighted images (DWIs) are used as input to train the corresponding subnetworks, and the FreeSurfer-based parcellation is used as the ground truth. Incorporating evidential loss, the subnetworks produce voxel-wise evidence that can be further parameterized as Dirichlet distributions and output evidence-based uncertainty. The evidence-based uncertainty is used as a criterion for ensemble prediction maps, and the final uncertainty heatmap is calculated using the entropy of the averaged evidence.
}
  \label{FIG.2}
\end{figure}

The goal of the proposed EVENet method is to compute an anatomical parcellation directly from the input dMRI data while providing an uncertainty map highlighting image voxels with low prediction confidence (see an overview of the method in Figure~\ref{FIG.2}). We first compute five diffusion parameter maps from the diffusion-weighted images (DWIs), describing different microstructural properties of the brain (Section 2.1). Then, for each input parameter, an evidential learning subnetwork is used to perform parameter-specific parcellation prediction and uncertainty estimation (Section 2.2). Finally, evidence-based ensemble learning is performed to fuse the five parameter-specific subnetworks and compute the final parcellation and uncertainty map (Section 2.3). 

\subsection{Network Architecture}

The overall network architecture has two major components, a backbone network that is based on the DDParcel method \citep{zhang2023ddparcel} and an uncertainty estimation network (Sections 2.2 and 2.3). In brief, the backbone network comprises five subnetworks, where each subnetwork takes an individual dMRI parameter as the input and is trained with the T1w-based FreeSurfer (FS) parcellation to predict brain parcellation from each diffusion parameter map. The network input includes five parameter maps derived from the diffusion tensor imaging (DTI) model \citep{pierpaoli1996toward}, including: FA that quantifies the variability of diffusion across different orientations, MD that quantifies the overall magnitude of diffusion, and E1, E2, and E3 that are the three eigenvalues of the diffusion tensor. Each subnetwork uses the FastSurferCNN \citep{henschel2020fastsurfer} architecture that is designed to perform T1w-based FS parcellation using deep learning. Given that different diffusion parameters describe the tissue properties from varying perspectives, resulting in parcellations differently from each other, evidential deep learning is used for estimation of the uncertainty of the outputs from the different subnetworks. To do so, the final classifier module with 1×1 convolution and softmax activation used in DDParcel is replaced with a softplus layer to enable a later integration of the evidence learning module that is described in detail below. 

\subsection{Evidential Deep Learning }
In the proposed EVENet, to enable the use of evidential learning, the output layer of each subnetwork includes the evidence learning module \citep{sensoy2018evidential,zou2022tbrats}], rather than using the classifier module from FastSurferCNN. Instead of applying a softmax on the outputs of the network, the outputs are identified as evidence $e_m^{q,c}$ in each subnetwork q that assigns the voxel m to the class n. The evidence learning module, as described in \citep{sensoy2018evidential}, transforms the evidence into beliefs and uncertainty. Beliefs $p_m^{q,c}$ are defined as:
\begin{equation}
    p_m^{q,c}=\frac{e_m^{q,c}}{s_m}
\end{equation}

where $S_{m}=\sum_{n=1}^{N} ( e_{m}^{q,n} + 1)$. For subnetwork q, at voxel m, the sum of beliefs $p_m^{q,n}$ across labels N and uncertainty $u_m^q$is 1 \citep{sensoy2018evidential}:
\begin{equation}
    \sum\limits_{n=1}^Np_m^{q,n}+u_m^q = 1
\end{equation}

The uncertainty of classification of subnetwork q at voxel m can be then directly obtained as:
\begin{equation}
    u_m^q=\frac{S_m-\sum_{n=1}^Ne_m^{q,n}}{S_m}=\frac N{S_m}
\end{equation}

The evidence and model parameters are learned by the minimization of the overall loss:
\begin{equation}
    L=L_{Dice}+\lambda L_{EDL}.
\end{equation}

Here, $L_{Dice}$ is the standard Dice loss function \citep{sudre2017generalised} computed against the ground truth T1w-based FS parcellation, which ensures stable parcellation across all FS regions of the entire brain. $L_{EDL}$ is the evidential loss function, which enables the network to learn the evidence $e_m^{n,q}$. The values of e are non-negative, ranging from zero to infinity, with larger values indicating higher confidence in the prediction. Specifically, $L_{EDL}$ is given as
\begin{equation}
    L_{EDL}=L_{rce}+\lambda_{kl}L_{KL}
\end{equation}

Here, $L_{rce}$ is a revised cross-entropy loss that adds Bayes risk to the conventional cross-entropy term to regulate the evidential learning process \citep{sensoy2018evidential}, and $L_{KL}$ is the Kullback-Leibler (KL) divergence that ensures incorrect labels produce less evidence \citep{sensoy2018evidential}. Specifically, the computation of $L_{rce}$ and $L_{KL}$ begins with using the evidence as the parameters of a Dirichlet distribution with $\alpha_{m}^{n,q}=e_{m}^{n,q}+1$, leading to the loss function \citep{sensoy2018evidential}: 
\begin{equation}
    L_{rce}=\sum\limits_{n=1}^{N}y_{m}^{n}(\psi(S_{m}) - \psi(\alpha_{m}^{n}))
\end{equation}

where $y_{m}^{n}$ is either one or zero for the ground truth class n in voxel m, and $\psi(.)$ is the digamma function. $L_{KL}$ is the Kullback-Leibler (KL) divergence, which is calculated as:
\begin{equation}
    \mathrm{L_{KL}}=\log\left(\frac{\Gamma\left(\Sigma_{n=1}^{N} \widetilde{\alpha}_{m}^{n}\right)}{\Gamma(N)\sum_{n=1}^{N} \Gamma\left(\widetilde{\alpha}_{m}^{n}\right)}\right)+\sum_{n=1}^{N}\left(\widetilde{\alpha}_{m}^{n}-1\right)\left[\psi\left(\widetilde{\alpha}_{m}^{n}\right)-\psi\left(\sum_{n=1}^{N}\widetilde{\alpha}_{m}^{n}\right)\right]
\end{equation}

where $\tilde{\boldsymbol{\alpha}}_m^n=y_m^n+(1-y_m^n)\odot\boldsymbol{\alpha}_m^n$. $\lambda$ and $\lambda_{kl}$are parameters that scale the contribution of the three loss functions.

\subsection{Evidence-based Ensemble Learning for Uncertainty Estimation}
Leveraging the outputs obtained from the five sub-networks, the entropy based on the averaged beliefs is used for the final uncertainty estimation. Specifically, each subnetwork produces a set of beliefs $p_m^{q,c}$ and an associated uncertainty $u_m^q$ for each voxel $m$, where $q$ indexes the subnetwork corresponding to each dMRI parameter, and $c$ indexes the class labels. To combine the information from the multiple subnetworks, we first compute an average belief across all subnetworks for each class at each voxel:
\begin{equation} 
    p_m^{'c} = \frac{1}{M} \sum_{q=1}^M p_m^{q,c} 
\end{equation}
where $N$ is the total number of class labels. The entropy $u'_m$ quantifies the uncertainty in the final prediction at voxel $m$, with higher entropy indicating higher uncertainty.
Next, we compute the entropy of the averaged belief distribution at each voxel to obtain the final uncertainty estimation:
\begin{equation}
    u'_m=-\sum_{c=1}^N{p_m^{'c}log(p_m^{'c})}
\end{equation}
where $N$ is the total number of class labels. The entropy $u'_m$ quantifies the uncertainty in the final prediction at voxel $m$, with higher entropy indicating higher uncertainty.

The entropy of classification probabilities is usually considered as an effective estimation of uncertainty \citep{luo2019applicability}, \citep{gawlikowski2023survey}. Here we used the entropy of beliefs to calculate the uncertainty, as the beliefs and probabilities have similar ranges and implications. The network-specific uncertainty is not directly related to the final uncertainty but can show how uncertain a subnetwork or a type of input is in general (See Figures \ref{FIG.4} and \ref{FIG.5}: some individual network uncertainty outputs are obviously brighter and clearer than the others), which also provides valuable information.

Finally, the evidence-based ensemble classification is performed based on the outputs of the five subnetworks. For each voxel, the subnetwork prediction with the minimum subnetwork uncertainty is adopted as the final prediction, described as:
\begin{equation}
    Q_m = argmin_{q\in\{FA,MD,E1,E2,E3\}}u_m^q
\end{equation}

\begin{equation}
    Class_m=argmax_{y\in\{1,2,3,..,N-1,N\}}p_m^{,Q_m,y}
\end{equation}

This means that for each voxel, we rely on the subnetwork that is most confident (i.e., has the lowest uncertainty) in its prediction at that location. The rationale behind this choice is that different dMRI parameters may provide more reliable information in different regions due to their sensitivity to different microstructural properties. By selecting the prediction from the subnetwork with the lowest uncertainty, we aim to leverage the most trustworthy information available at each voxel.

Entropy-based uncertainty affects the final segmentation by providing a measure of confidence in the sub-network predictions, which can be used to inform further processing or decision-making. For instance, regions with high entropy may correspond to areas where the model is uncertain due to factors such as image noise, anatomical variability, or pathology. By identifying these regions or modalities, clinicians or researchers can focus their attention on areas that may require manual review or additional imaging.

This approach is straightforward yet effective in that it leverages the evidence and uncertainty the model learns from each of the dMRI parameters, but it does not require complicated training. Since each dMRI parameter reveals different brain microstructures, it is also reasonable to treat each subnetwork as an individual source of evidence, but not to concatenate them into a trainable neural network again.

\subsection{Implementation and Parameter Settings}
Our method is implemented using Pytorch 2.0 \citep{paszke2019pytorch} and trained on a server equipped with NVIDIA RTX3090 GPUs. For each subnetwork, the development of our model is based on FastSurferCNN (\url{https://github.com/Deep-MI/FastSurfer}) and DDparcel (\url{https://github.com/zhangfanmark/DDParcel}). Code pieces in \citep{zou2022tbrats} are also used for the calculation of evidential loss (\url{https://github.com/Cocofeat/TBraTS}). We adopted the configurations such as learning rate and number of input channels as recommended by the FastSurferCNN paper, considering these values as a starting point for our model. We employed Adam as the optimizer \citep{kingma2014adam} because of its effectiveness in handling sparse gradients on noisy problems. A learning rate decay strategy was implemented to enhance training dynamics, with an initial rate of 0.01, then decreased by 95\% every five epochs over 200 epochs. The models were iteratively updated using a batch size of 8, which balances the trade-off between memory constraints and the benefits of mini-batch training. For the overall loss, the parameters  and kl are empirically set to 0.7 and 0.4, respectively, to balance the model’s focus between parcellation performance and quality of uncertainty estimation. To improve the generalization, we applied data augmentation techniques during training. Specifically, we performed random rotations (±15 degrees), scaling (90\% to 110\%), and horizontal flipping on the input images. These augmentations were applied on-the-fly during training. The use of data augmentation resulted in a 2\% increase in the Dice coefficient on the validation set, indicating enhanced model robustness.

\section{Experimental Evaluation}

\subsection{Experimental Datasets}
We evaluate the proposed EVENet method using dMRI datasets from multiple independently acquired populations (see Table 1), including: (1) 350 young healthy adults (28.1 ± 3.2 years old, 179 females and 171 males) in the Human Connectome Project (HCP) database \citep{glasser2013minimal,glasser2016multi}; (2) 50 young adults with diverse psychiatric conditions (37.6 ± 9.2 years old, 20 females and 30 males) consisting of 10 schizophrenia, 10 bipolar disorder, and 10 attention-deficit/hyperactivity disorder, and 20 healthy controls from the Consortium for Neuropsychiatric Phenomics (CNP) database \citep{poldrack2016phenome}; (3) 50 elderly adults (62.8 ± 7.1 years old, 25 females and 25 males), consisting of 25 healthy controls and 25 patients diagnosed with Parkinson's disease from The Parkinson’s Progression Markers Initiative (PPMI) database; (4) 11 cerebral small vessel disease patients (62.6 ± 19.8 years old, 2 females and 9 males) with visible white matter hyperintensity (WMH) at the Second Affiliated Hospital ofZhejiang University School of Medicine, China; (5) 22 neurosurgical patients (48.1 ± 23.1 years old, 9 females and 13 males) diagnosed with brain tumors (BT) from the First Affiliated Hospital of Sun Yat-sen University, China. In the rest of the paper, we refer to these datasets as the HCP, CNP, PPMI, WMH, and BT datasets. Usage of the in-house WMH and BT datasets was approved by the local ethics committees at the Second Affiliated Hospital of Zhejiang University School of Medicine and the First Affiliated Hospital of Sun Yat-sen University, respectively. 

In our study, the high-quality HCP datasets are used to train the brain parcellation model (n=200), as well as for model validation (n=100) and testing (n=50). The CNP, PPMI, WMH, and BT datasets, which are dMRI data collected through clinical acquisition protocols for studies focused on clinical applications, are utilized to evaluate the generalization capabilities of the trained model across various populations, acquisition protocols, and scanners. 

\subsubsection{MRI Acquisition and Preprocessing}
Table \ref{tab:demographic_dMRI_details} gives an overview of the diffusion image acquisitions of the datasets under study. These dMRI datasets were scanned with different diffusion imaging protocols, as follows. (1) The HCP data were acquired with a high-quality image acquisition protocol using a customized 3T Connectome Siemens Skyra scanner. The acquisition parameters are TE = 89.5 ms, TR = 5520 ms, voxel size = 1.25 × 1.25 × 1.25 mm$^3$. A total of 288 images were acquired for each subject, including 18 baseline images and 270 diffusion-weighted images evenly distributed at three shells of b = 1000/2000/3000 s/mm$^2$. (2) The CNP data were acquired using a 3T Siemens TrioTim scanner. For dMRI data, the acquisition parameters were: TE = 93 ms, TR = 9000 ms, and voxel size = 2 × 2 × 2 mm$^3$. Each subject underwent the acquisition of 65 volumes, consisting of 1 baseline image and 64 diffusion-weighted images at b = 1000 s/mm$^2$. For anatomical T1w data, the acquisition parameters included TE = 2.26 ms, TR = 1900 ms, and voxel size of 1 × 1 × 1mm$^3$. (3) The PPMI data were acquired using a 3T Siemens Trio scanner. The dMRI data acquisition parameters are TE = 88 ms, TR=7600 ms, and voxel size = 2 × 2 × 2 mm$^3$. A total of 65 volumes were acquired for each subject, including 1 baseline image with b = 0 s/mm$^2$ and 64 volumes at b = 1000 s/mm$^2$. (4) The WMH data were acquired using a 3T GE Healthcare MR750 scanner. For dMRI data, the acquisition parameters were: TE = 80.8 ms, TR = 8000 ms, and voxel size = 2 × 2 × 2 $^3$. The dMRI data was acquired with 30 non-collinear diffusion sensitization directions using a b-value of 1000 s/mm$^2$. Additionally, 5 volumes were obtained with no diffusion weighting (b-value = 0 s/mm$^2$). The other dMRI parameters included a flip angle of 90 degree and a slice thickness of 2 mm with no inter-slice gap. (5) The BT data were acquired using a 3T Siemens Prisma scanner. The acquisition parameters for the dMRI data included TE = 79 ms, TR = 22,000 ms, and a voxel size of 2 × 2 × 2 mm$^3$. A total of 65 volumes were acquired for each subject, including 1 baseline image with b = 0 s/mm$^2$ and 64 volumes with b = 1000 s/mm$^2$. 

For HCP, the provided dMRI data was processed following the HCP minimum processing pipeline \citep{glasser2013minimal}, including brain masking, motion correction, eddy current correction, EPI distortion correction, and rigid registration to the MNI space. For CNP, PPMI, WMH and BT, the dMRI data was processed as described in our previous study using a well-established pipeline \citep{zhang2018anatomically} (\url{https://github.com/pnlbwh/pnlpipe}), including eddy current-induced distortion correction, motion correction, and echo-planar imaging EPI distortion correction. Input diffusion parameters are computed using SlicerDMRI \citep{norton2017slicerdmri,Zhang2020-hc}. T1w-based FS parcellation is computed for each testing subject and used as ground truth for quantitative evaluation. In the BT dataset, two subjects fail to run parcellation using the Freesurfer software because of abnormal brain structures. Therefore, a subset of the BT dataset containing 20 subjects is used for quantitative comparison and the remaining 2 subjects are used for visual assessment of the uncertainty estimation results.

\begin{table}[h!]
\begin{threeparttable}
\caption{Demographic information and diffusion MRI acquisition details of the datasets under study\tnote{$a$}}
\label{tab:demographic_dMRI_details}
\setlength{\tabcolsep}{4pt}%

\begin{tabularx}{\textwidth}{@{}lcccp{3.5cm} >{\raggedright\arraybackslash}p{6cm}@{}}
\toprule
\textbf{Dataset} & \textbf{\# Subjects} & \textbf{Age}  & \textbf{Gender} & \textbf{Health Condition} & \textbf{dMRI data} \\
\midrule
HCP   & 350 & 28.1 ± 3.2 & 179 F, 171 M & 350 healthy & b= 1000 s/mm$^2$, 108 directions, TE/TR=89.5/5520 ms, resolution=1.25×1.25×1.25 mm$^3$ \\
CNP   & 50  & 37.6 ± 9.2 & 20 F, 30 M & 20 healthy, 10 BP, 10 SZ, 10 ADHD & b= 1000 s/mm$^2$, 64 directions, TE/TR=93/9000 ms, resolution=2×2×2 mm$^3$ \\
PPMI  & 50  & 62.8 ± 7.1 & 25 F, 25 M & 25 healthy, 25 PD & b= 1000 s/mm$^2$, 64 directions, TE/TR=88/7600 ms, resolution=2×2×2 mm$^3$ \\
WMH   & 11  & 62.6 ± 19.8 & 2 F, 9 M & 11 WMH & b= 1000 s/mm$^2$, 30 directions, TE/TR=80.8/8000 ms, resolution=2×2×2 mm$^3$ \\
BT    & 22  & 48.1 ± 23.1 & 9 F, 13 M & 22 BT & b= 1000 s/mm$^2$, 64 directions, TE/TR=79/22000 ms, resolution=2×2×2 mm$^3$ \\
\midrule
Total & \multicolumn{5}{c}{483} \\
\bottomrule
\end{tabularx}

\begin{tablenotes}[flushleft]\footnotesize
\item[${a}$] Abbreviations: Dataset: HCP - Human Connectome Project; CNP - Consortium for Neuropsychiatric Phenomics; PPMI - Parkinson’s Progression Markers Initiative; WMH - white matter hyperintensity; BT - Brain Tumor. Gender: F - female; M - male. Health Condition: BP - bipolar disorder; SZ - schizophrenia; ADHD - attention-deficit/hyperactivity disorder; PD - Parkinson’s disease.
\end{tablenotes}
\end{threeparttable}
\end{table}

\subsection{Experimental Design}
We perform three experimental evaluations. First, we conduct an ablation study to assess the effectiveness of different ensemble criteria within our parcellation framework (Section 3.2.1). Second, we compare the proposed method to five state-of-the-art dMRI parcellation methods (Section 3.2.2). Third, we evaluate the versatility and robustness of our method on multiple dMRI datasets with varying imaging conditions (Section 3.2.3).

\subsubsection{Ablation Study}
First, the effectiveness of the ensemble criteria and backbone network is evaluated in an ablation study, a comparison was performed among the following methods, including: (1) the probability-based method, (2) the entropy-based method, and (3)  the evidence-based method (proposed). Specifically, the probability-based method averages prediction probabilities derived from the original softmax activation layers. The entropy-based method calculates the entropy of the output probabilities and selects the label with minimal entropy as the final prediction for each voxel. The evidence-based method (proposed) employs the evidence learning module and conducts ensemble segmentation as described in Section 2.3. For the former two methods, we use the default loss settings recommended in the original papers to train the subnetworks. For evidence-based method, we also incorporate evidential deep learning loss. We test these three criteria of the ensemble on three backbone networks, including FastSurfer, nnU-Net, and Swin UNETR. The evaluation is performed using the HCP testing data. For each compared method and each HCP testing subject, the Dice score, recall, and intersection over union (IoU) for each FS region is computed between the prediction and the ground truth parcellation, and then the average score across all regions is obtained.

\subsubsection{Comparison to State-of-the-art Methods}
We then compare the proposed EVENet with several state-of-the-art methods in the literature, including FastSurfer \citep{henschel2020fastsurfer}, Swin UNETR \citep{hatamizadeh2021swin}, nnU-Net \citep{hatamizadeh2021swin}, \citep{isensee2021nnu}, and DDParcel \citep{zhang2023ddparcel}. The details for each compared method are introduced as follows. (1) FastSurfer is a method designed for predicting FS parcellations from structural MRI scans that provides efficient and accurate cortical reconstruction. To extend its applicability to diffusion dMRI data, we adapt FastSurfer to accept an individual dMRI parameter as input (FA is used as it yields the best parcellation performance across all the diffusion maps), enabling direct comparison with other methods. (2) Swin UNETR, initially designed for brain tumor segmentation in MRI images, reformulated image segmentation as a sequence to sequence prediction problem wherein multi-modal input data is projected into a 1D sequence of embedding and used as an input to a hierarchical Swin transformer as the encoder. Leveraging a hierarchical Swin Transformer as the encoder, Swin UNETR excels at handling multi-modal input data. In our study, we train a brain parcellation model with Swin UNETR using the same input images as our proposed method and evaluated its performance in predicting FS parcellations. (3) nnU-Net has earned acclaim for its adaptability and robust performance in medical image parcellation tasks. Leveraging the U-Net architecture, nnU-Net automatically configures a customized segmentation pipeline based on dataset characteristics and available hardware resources. In our study, we train a brain parcellation model with nnU-Net utilizing the same  multi-channel input images as our proposed method, and generate FS parcellations for comparative evaluation. (4) DDParcel is a recently proposed brain parcellation method directly using dMRI data. Unlike conventional approaches requiring inter-modality registration, DDParcel circumvents this need, thus minimizing parcellation errors arising from distortion artifacts and low dMRI image resolution. Designed specifically for dMRI data with its unique multi-parameter characteristics, DDParcel utilized a multi-level fusion strategy to harness information from the various dMRI parameters as inputs for training. Here, the comparison is performed using the HCP testing dataset. Evaluation metrics including Dice score, recall, and IoU were computed for each FS region, with statistical analyses performed to assess significant differences among the methods. This comprehensive evaluation framework ensured robust comparisons and provided insights into the strengths and limitations of each method for FS parcellation tasks.

\subsubsection{Comparison of Parcellation Performance on Different Datasets}
Furthermore, we evaluate EVENet on multiple datasets (including HCP, CNP, PPMI, WMH, and BT) to demonstrate the generalizability and robustness of our approach with varying imaging conditions. For each dataset, we first quantitatively compare our method with each of the state-of-the-art methods (see Section 3.2.2). The FreeSurfer-based parcellation is used as ground truth parcellation from which the Dice score between the predicted and ground truth parcellation maps is computed. It is worth noting that the FS parcellation failed for two subjects from the BT dataset due to the presence of the disease structures. For these two subjects, there is then no ground truth available for calculating the parcellation metrics. Therefore, a subset of the BT dataset containing 20 subjects is used to compute the metrics. Following that, visual comparisons of parcellation performance are provided to illustrate the practical effectiveness and reliability of our segmentation results. Finally, we evaluate the effectiveness of uncertainty estimation on the WMH and BT datasets. The manually segmented abnormal lesion regions are compared with the uncertainty heatmap. Through these experiments, we aim to establish the proposed method's effectiveness in accuracy, generalizability, and clinical utility.

\section{Experimental Results}

\subsection{Ablation Study}
Table \ref{tab:ablation_study} presents the ablation study results, showing that the evidence-based ensemble consistently outperforms the others in terms of Dice scores, recall, and IoU. The consistent superiority of the evidence-based ensemble indicates that it can more effectively capture target structures and preserve spatial relationships through its nuanced integration of voxel-level uncertainties. When considering the backbone networks, it is important to note the impact they have on overall performance. FastSurferCNN stands out for notably improving performance, suggesting its compatibility with ensemble methods. This compatibility implies that the ensemble approach leverages the strengths of FastSurferCNN effectively. On the other hand, nnU-Net exhibits minimal variation across different ensemble criteria. This could be attributed to its self-configured architectures, which are inherently optimized for performance. Consequently, it shows limited gains from further ensemble enhancements.

\begin{table}[h!]
\begin{threeparttable}
\caption{Ablation study with comparison to different ensemble criteria and backbone networks\tnote{$a$}}
\label{tab:ablation_study}
\setlength{\tabcolsep}{5pt}%
\begin{tabularx}{\textwidth}{@{}lXXXXXXXXX@{}}
\toprule
\multirow{3}{*}{\textbf{Backbone Network}} & \multicolumn{9}{c}{\textbf{Ensemble Criteria}} \\
\cmidrule(lr){2-10}
& \multicolumn{3}{c}{Probability-based} & \multicolumn{3}{c}{Entropy-based} & \multicolumn{3}{c}{Evidence-based} \\
\cmidrule(lr){2-4} \cmidrule(lr){5-7} \cmidrule(lr){8-10}
& Dice & Recall & IoU & Dice & Recall & IoU & Dice & Recall & IoU \\
\midrule
nnU-Net         & 0.751 & 0.791 & 0.661 & 0.755 & 0.790 & 0.668 & 0.758 & 0.794 & 0.673 \\
Swin UNETR      & 0.735 & 0.783 & 0.644 & 0.746 & 0.788 & 0.660 & 0.761 & 0.786 & 0.665 \\
FastSurferCNN   & 0.746 & 0.780 & 0.669 & 0.764 & 0.792 & 0.672 & \textbf{0.789} & \textbf{0.804} & \textbf{0.682}\\
\bottomrule
\end{tabularx}
\begin{tablenotes}[flushleft]\footnotesize
\item[${a}$] The table shows the performance of different backbone networks across various ensemble criteria.
\end{tablenotes}
\end{threeparttable}
\end{table}

\subsection{Comparison to State-of-the-art Methods}
Table \ref{tab:parcellation_performance} gives the comparison results of our proposed EVENet method compared with the FastSurfer, Swin UNETR, nnU-Net, and DDParcel methods. The results show that our proposed method obtains the highest Dice scores, followed by the DDParcel, nnU-Net, FastSurfer, and then the Swin UNETR method. Our proposed method also achieves the highest Mean Recall and Mean IoU scores among all methods evaluated, demonstrating improvement in parcellation benchmarks.

\begin{table}[h!]
\begin{threeparttable}
\caption{Parcellation performance in comparison with state-of-the-art methods\tnote{$a$}}
\label{tab:parcellation_performance}
\setlength{\tabcolsep}{10pt}%
\begin{tabularx}{\textwidth}{@{}lXXX@{}}
\toprule
\textbf{Model} & \textbf{Dice Score} & \textbf{Mean Recall} & \textbf{Mean IoU} \\
\midrule
FastSurfer   & 0.739 ± 0.071 & 0.792 ± 0.019 & 0.646 ± 0.011 \\
Swin UNETR   & 0.731 ± 0.092 & 0.784 ± 0.015 & 0.633 ± 0.006 \\
nnU-Net      & 0.749 ± 0.084 & 0.765 ± 0.011 & 0.663 ± 0.009 \\
DDParcel     & 0.770 ± 0.073 & 0.793 ± 0.015 & 0.670 ± 0.013 \\
\textbf{EVENet}      & \textbf{0.789} ± 0.061 & \textbf{0.804} ± 0.018 & \textbf{0.682} ± 0.010 \\
\bottomrule
\end{tabularx}
\begin{tablenotes}[flushleft]\footnotesize
\item[${a}$] For the three single-image input networks (FastSurfer, Swin UNETR, and nnU-Net), the network inputs are FA images, which produce the best parcellation results. Our proposed method and DDParcel allow inputs of multiple images and thus five and four dMRI parameters are used, respectively.
\end{tablenotes}
\end{threeparttable}
\end{table}

\subsection{Comparison of Parcellation Performance on Different Datasets}
The quantitative results of parcellation performance on different datasets are first studied. As shown in Table \ref{tab:parcellation_dice_scores}, our proposed method consistently outperforms the other methods across all datasets in terms of the Dice metric.

\begin{figure}[htbp]
  \centering
  \includegraphics[width=1.0\textwidth]{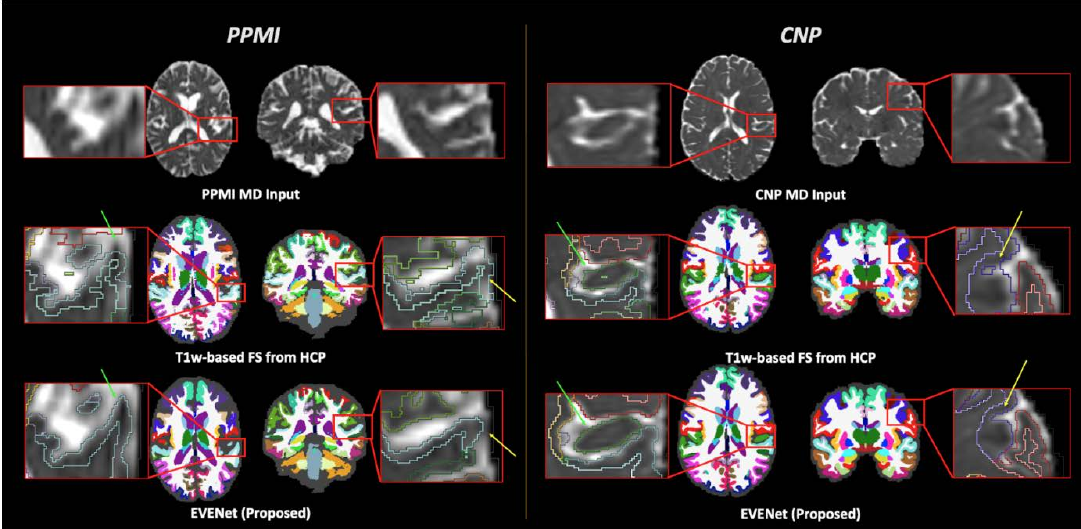}
  \caption{Visualization of parcellation results on randomly selected PPMI (left) and CNP (right) scans. The green and yellow arrows identify examples of tissue boundaries for easier comparison across parcellation labelmaps.
}
  \label{FIG.3}
\end{figure}

Figures \ref{FIG.2} and \ref{FIG.3} provide a visual comparison of the FS parcellation across the different methods in the HCP, CNP, and PPMI datasets. We can observe that our method generates a visually smoother segmentation that is more consistent with the tissue boundaries appearing on the input image. 

\begin{figure}[htbp]
  \centering
  \includegraphics[width=1.0\textwidth]{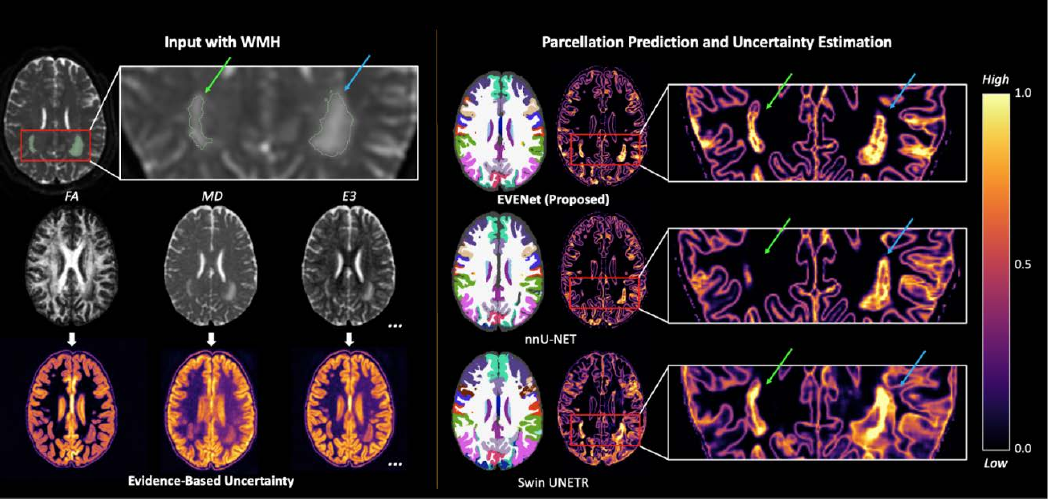}
  \caption{Visualization of uncertainty estimations on a randomly selected patient scan with WMH. The left part shows the input and the evidence-based uncertainty from three of the subnetworks with a comparison to the manually segmented WMH mask. The right side shows the final parcellation and uncertainty estimation. 
}
  \label{FIG.4}
\end{figure}

\begin{figure}[!ht]
  \centering
  \includegraphics[width=1.0\textwidth]{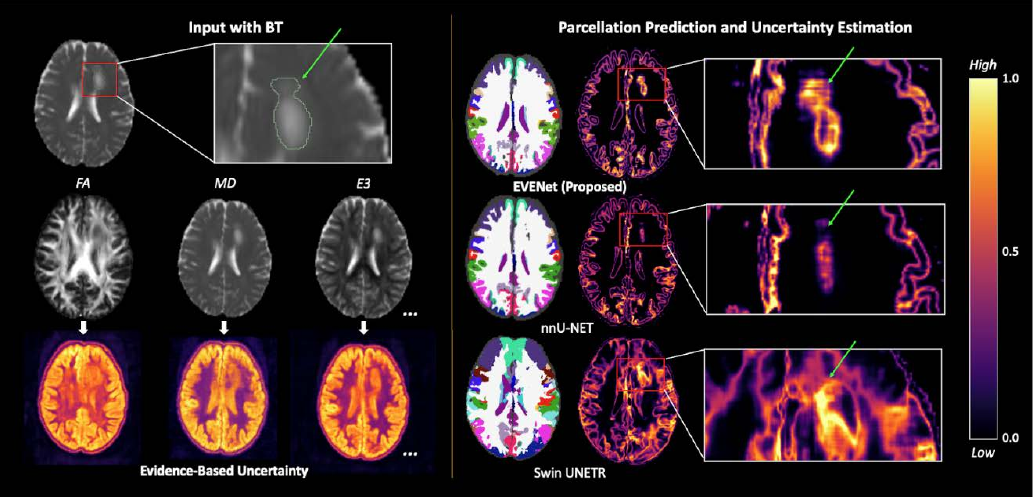}
  \caption{Visualization of uncertainty estimations on another randomly selected patient scan with BT. This scan fails to run properly on the FS software due to the existence of the tumor region. The left part shows the input and the evidence-based uncertainty from three of the subnetworks. The final output with a comparison to the manually segmented mask is provided. 
}
  \label{FIG.5}
\end{figure}

Figures \ref{FIG.4} and \ref{FIG.5} present a visualization of uncertainty estimations on randomly selected patient scans from the WMH and BT datasets. The left part displays the input alongside evidence-based uncertainty from three subnetworks. These heatmaps show rough views of uncertainty distribution, where high values are typically at tissue boundaries and pathological regions. Additionally, the final output is compared to the manually segmented lesion masks (white matter hyperintensity region or tumor). We can find that our network is capable of producing reasonable parcellation for unseen patient scans, and the final output uncertainty heatmaps clearly highlight the abnormal regions. Notably, the input with BT shown in Figure \ref{FIG.5} is one of the two subjects for whom FreeSurfer-based parcellation fails due to the presence of a large tumor region. In contrast, our method successfully performs parcellation while identifying  the abnormal lesion regions as regions of high uncertainty. More specifically, unlike the compared state-of-the-art methods (nn-Unet and Swin UNETR), in EVENet the parcellation of the gray matter remains accurate in the presence of a tumor or lesion.  In all the compared methods the lesion itself is not labeled as a lesion since a lesion label is not available in the training data. Nevertheless, EVENet outputs the uncertainty map that can be used for identification of abnormal lesion region, demonstrating the potential of EVENet in automatic segmentation of lesions.

\begin{table}[h!]
\begin{threeparttable}
\caption{Parcellation Dice Score on HCP, CNP, PPMI, BT, WMH datasets in comparison with state-of-the-art methods\tnote{$a$}}
\label{tab:parcellation_dice_scores}
\setlength{\tabcolsep}{8pt}%
\begin{tabularx}{\textwidth}{@{}lXXXXX@{}}
\toprule
\textbf{Dataset} & FastSurfer & Swin UNETR & nnU-Net & DDParcel & \textbf{EVENet} \\
\midrule
HCP   & 0.739 ± 0.071 & 0.731 ± 0.092 & 0.749 ± 0.084 & 0.770 ± 0.073 & \textbf{0.789 ± 0.061 }\\
CNP   & 0.653 ± 0.084 & 0.625 ± 0.102 & 0.673 ± 0.075 & 0.670 ± 0.094 & \textbf{0.693 ± 0.081 }\\
PPMI  & 0.669 ± 0.062 & 0.619 ± 0.087 & 0.680 ± 0.101 & 0.702 ± 0.074 & \textbf{0.709 ± 0.063} \\
BT    & 0.694 ± 0.077 & 0.685 ± 0.115 & 0.692 ± 0.081 & 0.713 ± 0.069 & \textbf{0.748 ± 0.053} \\
WMH   & 0.701 ± 0.095 & 0.671 ± 0.108 & 0.694 ± 0.105 & 0.747 ± 0.085 & \textbf{0.760 ± 0.092} \\
\bottomrule
\end{tabularx}
\begin{tablenotes}[flushleft]\footnotesize
\item[${a}$] The table compares the parcellation Dice scores across different datasets using various state-of-the-art methods.
\end{tablenotes}
\end{threeparttable}
\end{table}

\section{Discussion}
In this work, we introduced EVENet, an advanced Evidence-based Ensemble Neural Network for brain parcellation and uncertainty estimation using dMRI data. EVENet incorporates evidential deep learning with ensemble techniques to mitigate challenges such as handling out-of-distribution data, a common issue with conventional segmentation models. Our comprehensive evaluation demonstrated EVENet's superior performance in accuracy and robustness across diverse datasets, including those with pathological brain scans, underscoring its potential for significant impact in both research and clinical settings.

We showed that EVENet improved brain parcellation results both quantitatively and visually. Compared to the baseline and state-of-the-art FS parcellations, EVENet's results showed smoother tissue boundaries that better aligned with actual brain structures and also improved quantitative parcellation metrics. One of the main reasons is that EVENet is an ensemble approach that can utilize complementary information about brain microstructures extracted from the multiple dMRI parameters. Another reason is that EVENet uses evidential learning to estimate uncertainty in the model's predictions. This can help pinpoint potential areas in which the model is not accurate, identifying regions that may require further examination or refinement. For example, uncertainty heatmaps can reveal anatomical variations and inconsistencies in the data, highlighting areas where tissue boundaries are ambiguous or where partial volume effects are most pronounced due to the low resolution of dMRI data. 

Based on subjective logic, evidential deep learning offers an effective way of identifying out-of-distribution regions and cases. This evidence-based approach leverages the subjective logic theory to effectively manage the inherent uncertainties and complex information that dMRI offers \citep{jones2010twenty,le2003looking}. Unlike conventional models, evidential deep learning allows EVENet to express varying degrees of beliefs, accommodating the subtle and intricate diffusion patterns found in brain tissues. The overall uncertainty can be then easily derived from those beliefs, where unseen patterns in the training data will be highlighted. Those abnormal uncertainty heatmaps are valuable for identifying reliable areas/cases and those appearing to be out-of-distribution or needing further investigation in brain scans. Moreover, when different input parameters produce varied outcomes, the resulting uncertainty heatmaps from each subnetwork are also useful. They demonstrate the degree to which specific types or settings of inputs can inform us about brain structures, thereby aiding in making well-informed decisions and ongoing refinements.

EVENet’s robustness and generalizability were demonstrated on large testing datasets from multiple imaging sources and different populations. In the datasets from healthy controls and brain patient data without apparently abnormal brain regions (i.e., the CNP, and PPMI datasets), our method consistently outperformed the compared state-of-the-art deep learning methods. This strong generalization is essential for practical applications, ensuring reliable performance across diverse populations and varying imaging conditions. Ablation studies confirmed that the specific backbone structures and ensemble criteria used in EVENet’s subnetworks significantly contribute to its superior performance. When applied to out-of-distribution datasets with patients with abnormal brain lesions, EVENet introduced voxel-level uncertainty estimation, allowing for more precise identification of problematic areas in MRI scans. This capability can be important for clinical diagnostics. 

Furthermore, EVENet is designed to be a fast and compact tool for both parcellation and uncertainty estimation. The evidential component is integrated into the network structure, enabling uncertainty estimation without the need for retraining the model. On an RTX3090 GPU, EVENet completes both tasks in approximately 2 minutes. This efficiency makes EVENet particularly useful for guiding clinical assessments and ensuring reliable interpretations in a timely manner.

Potential limitations of the present study, including suggested future work to address limitations, are as follows. First, an interesting direction for future work is to integrate other imaging modalities such as T1-weighted and functional MRI into our framework. These modalities provide complementary structural and functional information that can further enhance the accuracy and robustness of uncertainty estimation. Multi-modal fusion strategies could be developed to leverage the strengths of each modality for a more comprehensive understanding of brain structure and function. Second, while our approach already offers a large computational advantage over the traditional registration-based approach, we acknowledge the potential for further optimization, particularly in the context of real-time processing for applications such as surgical planning and intraoperative guidance. Techniques such as model compression and pruning are promising directions to explore in the future, as they can reduce computational complexity. Although our current method meets the immediate demands of large-scale studies, further refinements in this area are feasible and will be considered in future work to enhance real-time capabilities. Third, recent advancements in self-supervised learning \citep{gui2024survey,huang2023self} and neural architecture search (NAS) \citep{kang2023neural,qin2023ng} have shown promise in medical imaging applications. Self-supervised learning could enable EVENet to leverage unlabeled data for pre-training, potentially improving performance on limited datasets. NAS could help in discovering more efficient network architectures tailored to our specific task, possibly reducing computational overhead. Incorporating these approaches in future work may enhance both the efficiency and generalizability of EVENet. Finally, we tested our method on populations with a wide age range including children, young adults, and elderly adults (see Supplementary Figure 1). However, our evaluation was also limited to a certain age range and did not extend to very young ages (e.g., babies and neonates). It is well known that the myelination of white matter in neonates is essentially different from the populations under study. Therefore, the curation of training data that reflects the anatomy of the specific populations might be needed to further improve the generalizability of EVENet across different age groups and developmental stages.

\section{Conclusion}
In conclusion, EVENet offers a robust and efficient solution for brain parcellation and uncertainty estimation using diffusion MRI data. By combining evidential deep learning with ensemble strategies, EVENet addresses key challenges in conventional segmentation models, providing superior accuracy and robustness across diverse datasets. The model's integration of multiple dMRI parameters and evidence-based ensemble methods enhances its performance, making it a valuable tool in both research and clinical contexts.

\section*{Data and Code Availability}

The data used in this project include the public HCP (\url{www.humanconnectome.org}), CNP (\url{https://openfmri.org/dataset/ds000030}), and PPMI (\url{http://www.ppmi-info.org}) datasets. The raw imaging data of the WMH and BT datasets are not publicly available because public availability would compromise participant confidentiality and participant privacy, but the derived diffusion MRI parameter maps will be made available upon request. The code and trained model will be made publically available at: \url{https://github.com/chenjun-li/EVENet}.

\section*{Acknowledgements}

This work is in part supported by the National Key R\&D Program of China (No. 2023YFE0118600), the National Natural Science Foundation of China (No. 62371107) and the National Institutes of Health (R01MH108574, P41EB015902, R01MH125860, R01MH119222, R01MH132610, R01NS125781, K99MH131850).

\bibliographystyle{unsrtnat}
\bibliography{references}  %%% Uncomment this line and comment out the ``thebibliography'' section below to use the external .bib file (using bibtex) .




% 	\bibitem{keshet2016prediction}
% 	Keshet, Renato, Alina Maor, and George Kour.
% 	\newblock Prediction-Based, Prioritized Market-Share Insight Extraction.
% 	\newblock In {\em Advanced Data Mining and Applications (ADMA), 2016 12th International 
%                       Conference of}, pages 81--94,2016.

% \end{thebibliography}

\end{document}